\documentstyle[preprint,aps]{revtex}
\begin{document}
\input{psfig}
\draft
\tighten
\title{
Structure analysis of the virtual Compton scattering amplitude at low energies
\thanks{This work was supported by the Deutsche Forschungsgemeinschaft
(SFB 201)}
}
\author{D.\ Drechsel,$^1$ G.\ Kn\"{o}chlein,$^1$ 
A.\ Yu.\ Korchin,$^{2,3}$ A.\ Metz,$^1$ and S.\ Scherer$^1$}
\address{$^1$ Institut f\"{u}r Kernphysik,
Johannes Gutenberg-Universit\"{a}t,
D-55099 Mainz, Germany}
\address{$^2$ Kernfysisch Versneller Instituut, 9747 AA Groningen, 
The Netherlands}
\address{$^3$ National Science Center ``Kharkov Institute of Physics
and Technology,'' 310108 Kharkov, Ukraine}
\date{November 5, 1997}
\maketitle
\begin{abstract}
   We analyze virtual Compton scattering off the nucleon at low
energies in a covariant, model-independent formalism.
   We define a set of invariant functions which, once the irregular nucleon
pole terms have been subtracted in a gauge-invariant fashion, is free 
of poles and kinematical zeros.
   The covariant treatment naturally allows one to implement the constraints
due to Lorentz and gauge invariance, crossing symmetry, and the
discrete symmetries.
   In particular, when applied to the $ep\to e'p'\gamma$ reaction, 
charge-conjugation symmetry in combination with nucleon crossing
generates four relations among the ten originally proposed generalized 
polarizabilities of the nucleon. 
\end{abstract}
\narrowtext

\section{Introduction}
\label{chapter_1}

The derivation of the structure of the general virtual Compton
scattering (VCS) amplitude from the nucleon has been a problem with a long
history \cite{Bardeen_68,Perrottet_73,Tarrach_75}.
The most sophisticated treatment of the general Compton process with 
both the initial and final state photons off shell, 
$\gamma^* N \rightarrow \gamma^* N$, was presented by Tarrach in 
Ref.\ \cite{Tarrach_75}. 
We will use this work as a starting point for our investigation of 
the low-energy VCS amplitude.

Whether one analyzes Compton scattering within the framework of a
given theoretical model or experimentally, it is in any case
desirable to perform the analysis in terms of a set of
amplitudes which is solely determined by the dynamics of the VCS
process with the kinematics being factored out. 
In the case of VCS with one or even two virtual photons, it is by no 
means trivial to find an adequate set of amplitudes which fulfills 
this requirement \cite{Perrottet_73,Tarrach_75}. 
In particular, as will become obvious in the following, it is a
central issue to construct tensor structures and corresponding amplitudes
which are free of poles or other kinematical constraints. 
This problem must be addressed on a model-independent level by taking
into account general symmetry principles like gauge and Lorentz
invariance and discrete symmetries like parity, time reversal, and 
charge conjugation.
We will discuss an ensemble of tensor structures and amplitudes with
the desired properties for the case of $\gamma^* N \rightarrow \gamma^* N$ 
which then can be applied to the case $\gamma^* N \rightarrow \gamma N$ 
with a real photon in the final state.
In particular, our results for the regular part of the VCS
amplitude can be expressed in terms of even fewer functions than
suggested in Ref.\ \cite{Tarrach_75}.

The process $\gamma^* N \rightarrow \gamma N$ will be analyzed at
the electron laboratories MAMI (Mainz), Jefferson Lab (Newport News), and
MIT-Bates by means of electron scattering off a proton target, $e p
\rightarrow e' p' \gamma$ \cite{experiments}. 
In the electron scattering process the genuine VCS amplitude interferes 
with the electron bremsstrahlung amplitude also known as the 
Bethe-Heitler process, which is completely determined by 
quantum electrodynamics and the electromagnetic form factors of 
the nucleon.
We will not discuss the Bethe-Heitler mechanism in this paper.

The scheduled VCS experiments have stimulated quite a few theoretical
activities.
Model-independent aspects of VCS have been studied in Refs.\  
\cite{Berg_61,Guichon_95,Scherer_96,Fearing_96,Drechsel_96}.
The various predictions for model-dependent quantities
related to VCS comprise the constituent quark model \cite{Guichon_95,Liu_96},
an effective Lagrangian approach \cite{Vanderhaegen_96}, calculations 
\cite{Korchin} in a coupled-channel unitary model \cite{Kneu},
field-theoretical models like the linear 
sigma model \cite{Metz_96,Metz_96a}, and heavy baryon chiral
perturbation theory (HBChPT) \cite{Hemmert_96a,Hemmert_96c}
as well as the Skyrme model \cite{Kim_97}. 
An overview of recent work on VCS may be found in Ref.\ \cite{VCSproc}.

The formalism applied most frequently to VCS at small final photon
energy and large momentum transfer has been developed in Ref.\ 
\cite{Guichon_95}.
In that work the regular part of the VCS amplitude has been 
parametrized in terms of ten generalized polarizabilities of the 
nucleon---three in the spin-independent and seven in the spin-dependent 
part of the amplitude.
Recently, a general proof has been given \cite{Drechsel_96} that
only two of the three generalized polarizabilities in the
spin-independent sector are independent of each other if
charge conjugation and nucleon crossing are applied. 
In the present work we will analyze the spin-dependent amplitudes on
the grounds of a covariant treatment.
The central result of our investigation will be that due to gauge 
invariance, Lorentz invariance, and invariance under parity, time reversal
and charge conjugation in combination with nucleon crossing the regular
part of the VCS amplitude can be written in terms of only six independent 
generalized polarizabilities instead of ten if one performs the same 
kinematical approximations as in Ref.\ \cite{Guichon_95}.

Our paper is organized as follows: In Sec.\ \ref{chapter_2} we briefly
review the formalism of Ref.\ \cite{Tarrach_75}, adapting the notation to our
conventions, and simplify the results according to our needs.
Part of the derivation is contained in Appendix \ref{appendix_a}.
In this section we will also specify the set of amplitudes we will
work with.
In Sec.\ \ref{chapter_3} we investigate the number of independent
generalized polarizabilities of the nucleon if one imposes the
same kinematical and symmetry constraints as in Ref.\ \cite{Guichon_95} but
in addition requires the VCS amplitude to be invariant under the
simultaneous transformation of charge conjugation and nucleon crossing.
Finally, we give a brief summary in Sec.\ \ref{chapter_4}.

\section{General Structure of the VCS Amplitude}
\label{chapter_2}

In this section we discuss the general form of the amplitude 
${\cal M}^{\gamma^{\ast}\gamma}$ for the VCS reaction 
$\gamma^{\ast} + N \to \gamma + N$. 
Before going into detail let us briefly explain our notation:
The initial (final) photon is characterized by the four-momentum 
$q^{\mu} = (\omega,\vec{q} \,)$ $[q'^{\mu} = (\omega',\vec{q}\,')]$, 
and the polarization vector 
$\varepsilon^{\mu} = (\varepsilon^0,\vec \varepsilon \,)$
$[\varepsilon'^{\mu} = (\varepsilon'^0,\vec \varepsilon \,')]$.
The four-momenta of the nucleons read $p_i^{\mu} = (E_i,\vec{p}_i )$,
$p_f^{\mu} = (E_f,\vec{p}_f)$.
For convenience, we introduce abbreviations for the sum of the photon
and the nucleon momenta,\footnote{We note that the definitions in 
Eq.\ (\ref{gl2_1}) differ by a factor of 2 from those used in Ref.\ 
\cite{Tarrach_75} but agree with Refs.\ \cite{Fearing_96}
and \cite{Drechsel_96}.}
\begin{equation} \label{gl2_1}
P = p_{i} + p_{f} \,, \quad Q = q + q' \, .
\end{equation}

The covariant result for ${\cal M}^{\gamma^{\ast}\gamma}$ turns out to be
a powerful tool for three reasons:
First of all, it can be used to investigate the number of independent
observables characterizing different kinematical approximations.
We study the consequences of the restriction to the lowest-order term 
in the real-photon energy $\omega'$ in order to determine the number 
of independent generalized polarizabilities.
Secondly, starting from the VCS results the transition to real Compton 
scattering (RCS) is simple and one is able to connect observables defined 
in RCS with those in VCS.
In particular, the relation between the third-order spin polarizabilities,
as defined by Ragusa \cite{Ragusa_93} for RCS, and the generalized
polarizabilities of Guichon {\em et al.} \cite{Guichon_95} can be obtained
\cite{Drechsel_97}.
Finally, our covariant result is appropriate to determine the general
form of the VCS amplitude in any specific frame.
In this paper, we only deal with the c.m. frame.

We start our analysis of the VCS amplitude considering the most general
case with two virtual photons.
The amplitude can be regarded as the contraction of the VCS tensor
$M^{\mu\nu}$ with the polarization vectors of the photons,
evaluated between the nucleon spinors in the initial and final states,
\begin{equation} \label{gl2_2}
{\mathcal{M}}^{\gamma^{\ast}\gamma^{\ast}} =
- i e^{2} \bar{u}(p_f,S_f) \varepsilon_{\mu} M^{\mu\nu}
\varepsilon_{\nu}^{\prime\ast} u(p_i,S_i)
\,.
\end{equation}
Throughout this paper we use the conventions of Bjorken and Drell
\cite{Bjorken_64}, where ${\mathcal{M}}^{\gamma^{\ast}\gamma^{\ast}}$
is the invariant matrix element of the VCS reaction.
The normalization of the nucleon spinor reads $\bar{u}(p,S) u(p,S) = 1$,
and we adopt Heaviside-Lorentz units where the square of the elementary 
charge is given by $e^{2}/4\pi \approx 1/137$.

   In order to disentangle new information from the VCS tensor,   
it is useful to separate from $M^{\mu\nu}$ the contribution
which is irregular in the limit $q \to 0$ or $q'\to 0$.
   For that purpose we divide $M^{\mu\nu}$ into a pole piece 
$M_{A}^{\mu\nu}$ and a residual part $M_{B}^{\mu\nu}$,
\begin{equation} \label{gl2_3}
M^{\mu\nu} = M_{A}^{\mu\nu} + M_{B}^{\mu\nu} \, .
\end{equation}
   In fact, such a splitting is not unique and we will follow the 
{\em convention} of Refs.\ \cite{Tarrach_75} and \cite{Guichon_95} of 
evaluating the $s$- und $u$-channel pole terms using electromagnetic
vertices of the form 
\begin{equation} \label{f1f2vertex}
\Gamma^\mu(p',p)=\gamma^\mu F_1(q^2)+i\frac{\sigma^{\mu\nu}q_\nu}{2M} F_2(q^2),
\,q=p'-p, \end{equation}
where $F_1$ and $F_2$ are the Dirac and Pauli form factors of the
proton, respectively.
   The explicit result for $M^{\mu\nu}_A$ is given in Eq.\ (18) of Ref.\  
\cite{Tarrach_75}.
   As a consequence of Low's theorem \cite{Low_58}, any calculation of pole 
terms involving on-shell equivalent forms of the nucleon electromagnetic
current yields the same irregular contribution to the VCS
matrix element (for a proof of this claim in the context of VCS, 
see Sec.\ IV B of Ref.\ \cite{Scherer_96}).
   It is advantageous to use the particular form of Eq.\ (\ref{f1f2vertex}),
since the resulting $M^{\mu\nu}_A$ separately satisfies all the 
symmetry requirements, in particular gauge invariance.
   Even though this terminology is not quite precise, we will adhere
to the common practice of referring to the $M^{\mu\nu}_A$ evaluated
with the vertices of Eq.\ (\ref{f1f2vertex}) as the ``Born terms.''
   The corresponding $M^{\mu\nu}_B$ will variously be denoted as the
regular or structure-dependent or residual or non-Born contribution.
   For a complete discussion of the ambiguity concerning what exactly is 
meant by ``Born terms,'' the interested reader is referred to Sec.\ IV of
Ref.\ \cite{Scherer_96}. 
   In the following, we are mainly interested in the non-Born contribution
to the Compton tensor, as this part {\em by definition} involves the 
generalized polarizabilities of Ref.\ \cite{Guichon_95} 
and the low-energy constants to be defined below.

   Using gauge invariance,
\begin{equation} \label{gl2_4}
q_{\mu} M^{\mu\nu} = q'_{\nu} M^{\mu\nu} = 0 \, ,
\end{equation}
a system of independent tensors serving as a basis of $M^{\mu\nu}$
was derived by Tarrach \cite{Tarrach_75}.
Once $M_{A}^{\mu\nu}$ and $M_{B}^{\mu\nu}$ are chosen to be gauge
invariant, we can construct both of them by use of the same basis
$M^{\mu\nu}$.

Since the work of Tarrach \cite{Tarrach_75} plays an important role in our
further analysis, we have summarized its results in Appendix \ref{appendix_a},
in particular the representation of the Compton tensor in terms of 18 
basis elements $T_{i}^{\mu\nu}$:
\begin{equation} \label{gl2_5}
M_{B}^{\mu \nu} = \sum_{i \in J}
 B_i (q^2 , q'^2 , q\cdot q' , q \cdot P) T_i^{\mu \nu} \,, \,\,\,
J = \left\{1,\ldots , 21 \right\} \backslash \left\{5,15,16 \right\}
\, .
\end{equation}
At this point we stress that the number of independent functions required 
for parametrizing the structure-dependent part is actually 18 instead of 
21 as suggested in Ref.\ \cite{Tarrach_75} (see Appendix \ref{appendix_a}).
The independent amplitudes $B_{i}$ are functions of four invariants
$q^2$, $q'^2$, $q \cdot q'$, and $q \cdot P$.
The kinematics of the general VCS process with on-shell nucleons is
completely specified by this set, and all other invariants 
can be expressed in terms of these variables.

So far we have considered both photons to be virtual.
We will now discuss the amplitude ${\mathcal{M}}^{\gamma^{\ast}\gamma}$ 
of the VCS process $\gamma^{\ast} + N \to \gamma + N$, with real photons
in the final state, i.e., $q'^{2} = 0$ and $\epsilon' \cdot q' = 0$.
In this specific case the tensors
$T_{3}^{\mu\nu}$, $T_{6}^{\mu\nu}$, and $T_{19}^{\mu\nu}$
do not contribute to the amplitude.
If we multiply the tensors
$\tilde{T}_{i}^{\mu\nu} \equiv T_{i}^{\mu\nu} (q'^{2} = 0)$
by the polarization vectors of both photons, we end up with
12 different structures which is the correct number of terms
\cite{Berg_61,Guichon_95,Scherer_96}.
As a consequence, the invariant VCS matrix element
${\mathcal{M}}_{B}^{\gamma^{\ast}\gamma} $ can be written as
\begin{equation} \label{gl2_10}
{\mathcal{M}}_{B}^{\gamma^{\ast}\gamma} =
 - i e^2
\bar{u} ( p_f,S_f) \sum_{i=1}^{12} \varepsilon_{\mu} \rho_i^{\mu\nu}
\varepsilon'^*_{\nu} f_i(q^2 , q \cdot q' , q \cdot P) u(p_i, S_i) \,.
\end{equation}
Equation (\ref{gl2_10}), together with the explicit results for the 
quantities $\epsilon_{\mu} \rho_{i}^{\mu\nu} \epsilon_{\nu}'^*$ in
Eq.\ (\ref{gl2_8}) of Appendix A, defines the general structure of the 
VCS amplitude  with $q^2 \neq 0$ and $q'^2 = 0$.
Since Eq.\ (\ref{gl2_10}) is Lorentz invariant, it is frame independent, 
and it allows one to incorporate the constraints from the discrete symmetries 
in a rather simple way.

In the following we will work in the c.m.\ frame, i.e.,
\begin{equation} \label{gl2_11}
\vec p_i = - \vec q \, , \quad
\vec p_f = - \vec q{\,'} \, ,
\end{equation}
and will use an orthonormal basis defined by the momenta of the photons,
\begin{equation} \label{gl2_12}
\hat{e}_z = \hat{q} \, , \quad
\hat{e}_y = \frac{\hat{q} \times \hat{q}'}{\sin \theta} \, , \quad
\hat{e}_x = \hat{e}_y \times \hat{e}_z \, ,
\end{equation}
with $\theta$ denoting the scattering angle between $\hat{q}$ and
$\hat{q}'$.

The matrix element can be decomposed into a transverse and a
longitudinal part,
\begin{equation} \label{gl2_13}
{\mathcal{M}}_{B}^{\gamma^{\ast} \gamma} = i e^2 \chi_f^{\dagger}
\left( \vec \varepsilon_T
\cdot \vec M_T + \frac{q^2}{\omega^2} \varepsilon_z M_z
\right)
\chi_i
\, ,
\end{equation}
where current conservation has been used,
\begin{equation} \label{gl2_14}
q_{\mu} \varepsilon^{\mu} = 0 \, , \quad
q_{\mu} M_{B}^{\mu\nu} = 0\, ,
\end{equation}
at the leptonic and the hadronic vertices, respectively.
Note that in the VCS process discussed in this paper the polarization 
vector of the initial photon is generated by the electromagnetic transition 
current of the electron,
$\varepsilon^{\mu} = e \bar{u}_{e'} \gamma^{\mu} u_{e} / q^{2}$. 
Current conservation allows one to perform the gauge transformation 
$\varepsilon^{\mu}
\rightarrow a^{\mu} = \varepsilon^{\mu} + \zeta q^{\mu}$. 
Then the choice $\zeta = - \vec \varepsilon \cdot \vec q/\omega^2$
leads to the polarization vector  
\begin{equation} \label{gl2_15}
a^{\mu} = \left(0,\vec \varepsilon_T + \frac{q^2}{\omega^2}
\varepsilon_z \hat{q} \right)
\end{equation}
and thereby 
results in the specific form of
${\mathcal{M}}_{B}^{\gamma^{\ast} \gamma}$ in Eq.\ (\ref{gl2_13}).

For the following discussion it is useful to decompose the VCS matrix element
in Pauli space. 
We choose the parametrization and the corresponding amplitudes defined 
in Ref.\ \cite{Hemmert_96a}.
The transverse and longitudinal matrix elements can, respectively,
be parametrized in terms of eight and four structures,
\begin{eqnarray} \label{gl2_16}
\vec \varepsilon_T \cdot \vec M_T & = & {\vec{\varepsilon}}\,'^* \cdot
{\vec{\varepsilon}}_T
A_1
+ {\vec{\varepsilon}}\,'^* \cdot {\hat{q}} {\vec{\varepsilon}}_T
\cdot {\hat{q}}'
A_2 \nonumber\\
& &+ i \vec \sigma \cdot \left( {\vec{\varepsilon}}\,'^*
\times {\vec{\varepsilon}}_T
 \right) A_3
+ i \vec \sigma \cdot \left( {\hat{q}}' \times {\hat{q}} \right)
{\vec{\varepsilon}}\,'^* \cdot {\vec{\varepsilon}}_T A_4 \nonumber \\
&&+ i \vec \sigma \cdot \left( {\vec{\varepsilon}}\,'^*
\times {\hat{q}} \right)
{\vec{\varepsilon}}_T \cdot {\hat{q}}' A_5
+ i \vec \sigma \cdot \left(
{\vec{\varepsilon}}\,'^* \times {\hat{q}}' \right)
{\vec{\varepsilon}}_T \cdot {\hat{q}}' A_6 \nonumber \\
&&- i \vec \sigma \cdot \left( {\vec{\varepsilon}}_T \times {\hat{q}}' \right)
{\vec{\varepsilon}}\,'^* \cdot {\hat{q}} A_7
- i \vec \sigma \cdot \left( {\vec{\varepsilon}}_T \times {\hat{q}} \right)
{\vec{\varepsilon}}\,'^* \cdot {\hat{q}} A_8 \, , \label{trans}
\\
\label{gl2_17}
M_z & = & {\vec{\varepsilon}}\,'^* \cdot {\hat{q}} A_9
+ i \vec \sigma \cdot \left( {\hat{q}}' \times {\hat{q}} \right)
{\vec{\varepsilon}}\,'^* \cdot {\hat{q}} A_{10}
\nonumber \\
& & + i \vec \sigma \cdot \left( {\vec{\varepsilon}}\,'^* \times {\hat{q}}
\right) A_{11}
+ i \vec \sigma \cdot \left( {\vec{\varepsilon}}\,'^* \times {\hat{q}}'
\right) A_{12} \,
. \label{long}
\end{eqnarray}

\section{Generalized Polarizabilities}
\label{chapter_3}
We now apply the general result of Eq.\ (\ref{gl2_10}), in order to 
determine the number of independent polarizabilities emerging from the
leading-order term of a consistent expansion of the residual 
amplitude ${\cal M}_{B}^{\gamma^{\ast} \gamma}$ in the energy $\omega'$
of the outgoing, real photon \cite{Guichon_95}.
For completeness we include the results of Ref.\ \cite{Drechsel_96} for the 
spin-independent polarizabilities in our presentation.

The definition of the generalized polarizabilities in VCS is based upon 
the multipole representation of 
${\mathcal{M}}_{B}^{\gamma^{\ast} \gamma}$ \cite{Guichon_95,Arenhoevel_74}. 
In Ref.\ \cite{Guichon_95} the multipoles 
$H^{(\rho' L' , \rho L)S}(\omega' , \bar{q})$ were introduced,
where $\rho \, (\rho')$ denotes the type of the initial (final) photon
($\rho = 0 : \;$ charge, C; $\rho = 1 : \;$ magnetic, M;
$\rho = 2 : \;$ electric, E). 
The initial (final) orbital angular momentum is characterized by $L \, (L')$,
and the quantum number $S$ distinguishes between non-spin-flip 
$(S = 0)$ and spin-flip $(S = 1)$ transitions.

According to the low-energy theorem for VCS \cite{Guichon_95,Scherer_96}, 
which is  an extension of the famous low-energy theorem for RCS derived 
by Low \cite{Low}, and Gell-Mann and Goldberger \cite{LET}, 
${\cal M}_{B}^{\gamma^{\ast} \gamma}$ is at least linear
in the energy of the real photon.
If one restricts oneself to the lowest-order term in $\omega'$, only 
electric and magnetic dipole radiation of the outgoing photon contributes
to the amplitude.
In that case selection rules for parity and angular momentum
allow for three scalar multipoles $(S = 0)$ and seven vector multipoles 
$(S = 1)$, leading to the same number of generalized polarizabilities
(see Ref.\ \cite{Guichon_95} for more details concerning the 
de\-finition of the generalized polarizabilities).

It turns out that multipoles containing an electric transition can be 
replaced by more appropriate definitions.
In the case of the outgoing photon only the leading term in
$\omega' = |\, \vec{q} \,' |$ is considered.
Therefore, Siegert's theorem \cite{Siegert_37}, together with the 
continuity equation, offers the possibility to express the electric 
transitions in terms of the charge transitions.
In contrast to the final state kinematics, one is interested in 
considering an arbitrary three-momentum $\bar{q}$ of the virtual photon
in the initial state, which allows for investigating the momentum 
dependence of the polarizabilities.
Accordingly, one has to be careful when replacing the electric multipoles 
in the initial state with charge multipoles, because the difference 
between electric and charge multipoles must not be neglected.
This leads to so-called mixed multipoles $\hat{H}^{(\rho' L' , L)S}$ 
\cite{Guichon_95}, which are no longer characterized by a well-defined 
multipole type of the incoming photon.

Bearing these considerations in mind, the generalized polarizabilities 
can be defined through
\begin{mathletters}
\label{gl3_1}
\begin{eqnarray} \label{gl3_1a}
P^{(\rho' L' , \rho L)S} (\bar{q}^2) & = &
\left[ \frac{1}{\omega'^{L} \bar{q}^{L}}
H^{(\rho' L' , \rho L)S} (\omega' , \bar{q}) \right]_{\omega' = 0}
\qquad (\rho , \rho' = 0,1) \, ,
\\ \label{gl3_1b}
\hat{P}^{(\rho' L' , L)S} (\bar{q}^2) & = &
\left[ \frac{1}{\omega'^{L} \bar{q}^{L+1}}
\hat{H}^{(\rho' L' , L)S} (\omega' , \bar{q}) \right]_{\omega' = 0}
\qquad (\rho' = 0,1) \, ,
\end{eqnarray}
\end{mathletters}
as functions of $\bar{q}^2$ \cite{Guichon_95}.
Contrary to multipoles containing an electric transition in the initial 
state, the multipoles in Eqs.\ (\ref{gl3_1a}) and (\ref{gl3_1b}) have a 
path-independent limit as $\bar{q},\omega' \to 0$.
In particular, in the $\omega'$-$\bar{q}$-plane the limits along the RCS line 
($\bar{q} = \omega'$) and along the VCS line ($\omega'=0$) coincide.
This behavior of the multipoles makes it possible to relate, 
at $\bar{q} = 0$, some of the corresponding generalized polarizabilities 
to the polarizabilities defined in RCS. 
An extended discussion on the low-energy behavior of the multipoles
and of the generalized polarizabilities can be found in 
Ref.\ \cite{Guichon_95}.

Two of the three scalar polarizabilities can be understood as
generalizations of the well-known electric $(\alpha)$ and
magnetic $(\beta)$ polarizabilities in RCS,
\begin{mathletters}
\begin{eqnarray} \label{gl3_2a}
\alpha (\bar{q}^2) & = & - \frac{e^{2}}{4 \pi} \sqrt{\frac{3}{2}}
P^{(01,01)0} (\bar{q}^2) \,,
\\ \label{gl3_2b}
\beta (\bar{q}^2) & = & - \frac{e^{2}}{4 \pi} \sqrt{\frac{3}{8}}
P^{(11,11)0} (\bar{q}^2) \,.
\end{eqnarray}
\end{mathletters}
To apply Eqs.\ (\ref{gl3_2a}) and (\ref{gl3_2b}) in Gaussian units one has to 
replace the factor $e^{2}/4\pi$ by $\alpha_{QED} = e^{2}_{Gauss}$.
This replacement ensures that the numerical numbers of $\alpha$ and $\beta$
in the Heaviside-Lorentz system and in the Gauss system are the same.
Note that by definition the generalized polarizabilities of Ref.\ 
\cite{Guichon_95} do not depend on the value of $e^{2}$.

Since we perform an expansion in $\omega'$, we will introduce two 
variables, 
\begin{mathletters}
\begin{eqnarray} \label{gl3_3a}
\omega_{0} & = & \omega |_{\omega' = 0} = M - E_i
= M - \sqrt{M^{2} + \bar{q}^{2}} \,,
\\ \label{gl3_3b}
Q_{0}^{2} & = & Q^{2} |_{\omega' = 0} = - q^{2} |_{\omega' = 0}
= - 2 M \omega_{0} \,.
\end{eqnarray}
\end{mathletters}
Following Guichon {\em et al.} \cite{Guichon_95}, the leading terms of
the amplitudes $A_{i}$ from Eqs.\ (\ref{gl2_16}) and (\ref{gl2_17}) read
\begin{mathletters}
\begin{eqnarray} \label{gl3_4a}
A_1 & = & \omega' \sqrt{\frac{E_i}{M}}
 \biggl[ - \sqrt{\frac{3}{2}} \omega_0 P^{(01,01)0}(\bar{q}^2)
 - \frac{3}{2} \bar{q}^2 \hat{P}^{(01,1)0}(\bar{q}^2)
 - \sqrt{\frac{3}{8}} \bar{q} \cos \theta P^{(11,11)0}(\bar{q}^2) \biggr]
\nonumber \\
& &  + {\mathcal{O}}(\omega'^{2}) \, ,
 \\ \label{gl3_4b}
A_2 & = & \omega' \sqrt{\frac{E_i}{M}}
 \biggl[ \sqrt{\frac{3}{8}} \bar{q} P^{(11,11)0}(\bar{q}^2) \biggr]
 + {\mathcal{O}}(\omega'^{2}) \, ,
 \\ \label{gl3_4c}
A_3 & = & \omega' \sqrt{\frac{E_i}{M}}
 \frac{3}{4} \biggl[ - 2 \omega_0 P^{(01,01)1}(\bar{q}^2)
 + \sqrt{2} \bar{q}^2 \Bigl [ P^{(01,12)1}(\bar{q}^2)
 - \sqrt{3} \hat{P}^{(01,1)1}(\bar{q}^2) \Bigr ]
 \nonumber\\
 & & \qquad \qquad + \Bigl ( - \bar{q} P^{(11,11)1}(\bar{q}^2)
 + \sqrt{\frac{3}{2}} \omega_0 \bar{q} P^{(11,02)1}(\bar{q}^2)
 + \sqrt{\frac{5}{2}} \bar{q}^3 \hat{P}^{(11,2)1}(\bar{q}^2) \Bigr )
 \cos \theta \biggr]
\nonumber \\
& & 
 + {\mathcal{O}}(\omega'^{2}) \, ,
 \\ \label{gl3_4d}
A_4 & = & \omega' \sqrt{\frac{E_i}{M}}
 \frac{3}{4} \biggl[ - \bar{q} P^{(11,11)1}(\bar{q}^2)
 - \sqrt{\frac{3}{2}} \omega_0 \bar{q} P^{(11,02)1}(\bar{q}^2)
 - \sqrt{\frac{5}{2}} \bar{q}^3 \hat{P}^{(11,2)1}(\bar{q}^2) \biggr]
 + {\mathcal{O}}(\omega'^{2}) \, ,
 \\ \label{gl3_4e}
A_5 & = & - A_4 \, , \vphantom{\frac{1}{1}}
 \\ \label{gl3_4f}
A_6 & = & {\mathcal{O}}(\omega'^{2}) \, ,  \vphantom{\frac{1}{1}}
 \\ \label{gl3_4g}
A_7 & = & \omega' \sqrt{\frac{E_i}{M}}
 \frac{3}{4} \biggl[ \bar{q} P^{(11,11)1}(\bar{q}^2)
 - \sqrt{\frac{3}{2}} \omega_0 \bar{q} P^{(11,02)1}(\bar{q}^2)
 - \sqrt{\frac{5}{2}} \bar{q}^3 \hat{P}^{(11,2)1}(\bar{q}^2) \biggr]
 + {\mathcal{O}}(\omega'^{2}) \, ,
 \\ \label{gl3_4h}
A_8 & = & \omega' \sqrt{\frac{E_i}{M}}
 \biggl[ - \frac{3}{\sqrt{2}} \bar{q}^2 P^{(01,12)1}(\bar{q}^2) \biggr]
 + {\mathcal{O}}(\omega'^{2}) \, ,
 \\ \label{gl3_4i}
A_9 & = & \omega' \sqrt{\frac{E_i}{M}}
 \biggl[ - \omega_0 \sqrt{\frac{3}{2}} P^{(01,01)0}(\bar{q}^2) \biggr]
 + {\mathcal{O}}(\omega'^{2}) \, ,
 \\ \label{gl3_4j}
A_{10} & = & \omega' \sqrt{\frac{E_i}{M}}
 \biggl[ - \frac{3 \sqrt{3}}{2 \sqrt{2}} \omega_0 \bar{q}
 P^{(11,02)1}(\bar{q}^2) \biggr]
 + {\mathcal{O}}(\omega'^{2}) \, ,
 \\ \label{gl3_4k}
A_{11} & = & \omega' \sqrt{\frac{E_i}{M}}
 \biggl[ - \frac{3}{2} \omega_0 P^{(01,01)1}(\bar{q}^2)
 + \frac{3 \sqrt{3}}{2 \sqrt{2}} \omega_0 \bar{q} \cos \theta
 P^{(11,02)1}(\bar{q}^2) \biggr]
 + {\mathcal{O}}(\omega'^{2}) \, ,
 \\ \label{gl3_4l}
A_{12} & = & \omega' \sqrt{\frac{E_i}{M}}
 \biggl [ \frac{\sqrt{3} \omega_0}{2 \bar{q}} \Bigl [ P^{(11,00)1}(\bar{q}^2)
 - \sqrt{2} \bar{q}^2 P^{(11,02)1}(\bar{q}^2) \Bigr ] \biggr]
 + {\mathcal{O}}(\omega'^{2}) \, .
\end{eqnarray}
\end{mathletters}
In the derivation we made use of the transformation (\ref{glb_3})
(see Appendix \ref{appendix_b}) between the $A_{i}$ and the amplitudes
defined in Ref.\ \cite{Guichon_95}.
We note that the relation between the matrix element $T^{VCS}$
in Ref.\ \cite{Guichon_95} and ${\mathcal{M}}_{B}^{\gamma^{\ast} \gamma}$ 
is given by
\begin{equation}
{\mathcal{M}}_{B}^{\gamma^{\ast} \gamma} = - i e^{2} T^{VCS}/2M \, .
\end{equation}

Another low-energy expansion of the amplitudes $A_{i}$ is obtained, if 
the covariant result of Eq.\ (\ref{gl2_10}) is evaluated in the c.m.\ frame.
Restricting ourselves to terms linear in $\omega'$, the expansion reads
\begin{mathletters}
\label{gl3_5}
\begin{eqnarray} \label{gl3_5a}
A_1 & = & \omega' \sqrt{\frac{E_i + M}{2 M}}
 \biggl[ - \omega_0 f_1 - 2 M \bar{q}^2 f_3 + 2 \omega_0 f_{10}
 + \Bigl ( \bar{q} f_1 + 2 M \omega_0 \bar{q} f_3
 - 2 \frac{\omega_0^2}{\bar{q}} f_{10} \Bigr ) \cos \theta \biggr]
\nonumber \\
& & 
 + {\mathcal{O}}(\omega'^{2}) \, ,
 \\ \label{gl3_5b}
A_2 & = & \omega' \sqrt{\frac{E_i + M}{2 M}}
 \biggl[ - \bar{q} f_1 - 2 M \omega_0 \bar{q} f_3
 + 2 \frac{\omega_0^2}{\bar{q}} f_{10} \biggr]
 + {\mathcal{O}}(\omega'^{2}) \, ,
 \\ \label{gl3_5c}
A_3 & = & \omega' \sqrt{\frac{E_i + M}{2 M}}
 \biggl[ - M \omega_0 f_5 + M \omega_0^2 f_8 - 2 \omega_0 f_{10}
 - M \omega_0^2 f_{12} \vphantom{\frac{1}{1}}
 \nonumber\\
 & & \qquad \qquad \qquad + \Bigl ( \frac{M \omega_0^2}{\bar{q}} f_5
 - M \omega_0 \bar{q} f_8 + 2 \bar{q} f_{10}
 + \frac{M \omega_0^3}{\bar{q}} f_{12} \Bigr ) \cos \theta
 \biggr]
 + {\mathcal{O}}(\omega'^{2}) \, ,
 \\ \label{gl3_5d}
A_4 & = & \omega' \sqrt{\frac{E_i + M}{2 M}}
 \biggl[ - 4 \frac{M \omega_0}{\bar{q}} f_{10} \biggr]
 + {\mathcal{O}}(\omega'^{2}) \, ,
 \\ \label{gl5_5e}
A_5 & = & - A_{4}  \vphantom{\frac{1}{1}} \, ,
 \\ \label{gl3_5f}
A_6 & = & {\mathcal{O}}(\omega'^{2})  \vphantom{\frac{1}{1}} \, ,
 \\ \label{gl3_5g}
A_7 & = & \omega' \sqrt{\frac{E_i + M}{2 M}}
 \biggl[ - \frac{M \omega_0^2}{\bar{q}} f_5 + M \omega_0 \bar{q} f_8
 - 2 \bar{q} f_{10} - \frac{M \omega_0^3}{\bar{q}} f_{12} \biggr]
 + {\mathcal{O}}(\omega'^{2}) \, ,
 \\ \label{gl3_5h}
A_8 & = & \omega' \sqrt{\frac{E_i + M}{2 M}}
 \biggl[ - 8 M^2 \omega_0 f_6 - M \omega_0 f_7 - M \omega_0^2 f_8
 - 4 M^2 \omega_0 f_9 + 2 \omega_0 f_{10}
 \nonumber\\
 & & \qquad \qquad \qquad - 4 M \omega_0 f_{11} + M \omega_0^2 f_{12}
 \biggr]
 + {\mathcal{O}}(\omega'^{2}) \, ,
 \\ \label{gl3_5i}
A_9 & = & \omega' \sqrt{\frac{E_i + M}{2 M}}
 \biggl[ - \omega_0 f_1 + 2 M \bar{q}^2 f_2 + 4 M \omega_0^2 f_6
 + 2 M \omega_0^2 f_9 - 2 M \omega_0^2 f_{12} \biggr]
 + {\mathcal{O}}(\omega'^{2}) \, ,
 \\ \label{gl3_5j}
A_{10} & = & \omega' \sqrt{\frac{E_i + M}{2 M}}
 \biggl[ 2 \omega_0 \bar{q} f_4 - \frac{\omega_0^3}{2\bar{q}} f_5
 - \frac{\omega_0 \bar{q}}{2} f_7 - 2 \frac{\omega_0^2}{\bar{q}} f_{10}
 - 2 \omega_0 \bar{q} f_{11} - \frac{M \omega_0^3}{\bar{q}} f_{12}
 \biggr]
\nonumber \\
& & 
 + {\mathcal{O}}(\omega'^{2}) \, ,
 \\ \label{gl3_5k}
A_{11} & = & \omega' \sqrt{\frac{E_i + M}{2 M}}
 \biggl[ 2 \bar{q}^2 f_4 - \frac{\omega_0^2}{2} f_5
 - \frac{\omega_0^2}{2} f_7 - 2 \omega_0 f_{10} - 2 \omega_0^2 f_{11}
 - 2 M \omega_0^2 f_{12}
 \nonumber\\
 & & \qquad \qquad \qquad + \Bigl ( - 2 \omega_0 \bar{q} f_4
 + \frac{\omega_0^3}{2 \bar{q}} f_5
 + \frac{\omega_0 \bar{q}}{2} f_7 + 2 \frac{\omega_0^2}{\bar{q}} f_{10}
 + 2 \omega_0 \bar{q} f_{11} + \frac{M \omega_0^3}{\bar{q}} f_{12} \Bigr )
 \cos \theta \biggr]
\nonumber \\
& & 
 + {\mathcal{O}}(\omega'^{2}) \, ,
 \\ \label{gl3_5l}
A_{12} & = & \omega' \sqrt{\frac{E_i + M}{2 M}}
 \biggl[ - \frac{M \omega_0^2}{\bar{q}} f_5 + \Bigl ( M \omega_0 \bar{q}
 - 2 \frac{M^2 \omega_0^2}{\bar{q}} \Bigr ) f_{12} \biggr]
 + {\mathcal{O}}(\omega'^{2}) \, .
\end{eqnarray}
\end{mathletters}
Because of the expansion in $\omega'$, the functions in Eqs.\ (\ref{gl3_5}) 
must have the arguments
$f_{i} = f_{i} |_{\omega' = 0} = f_{i}( - Q_{0}^{2} , 0 , 0)$.

Until now we have not used the transformation properties of the
functions $f_{i}$ with respect to photon crossing and the combination
of charge conjugation and nucleon crossing,
which can be obtained from Eqs.\ (\ref{gl2_6}) and (\ref{gl2_9}).
In particular, the behavior with respect to charge conjugation leads to the
conclusion that $f_{3}$, $f_{4}$, $f_{8}$, and $f_{10}$ are odd functions 
of $q \cdot P$ (see Eq.\ (\ref{gl2_6d})).
Consequently, they are at least linear in $\omega'$ and, therefore,
do not contribute to the leading-order terms
in Eqs.\ (\ref{gl3_5a})--(\ref{gl3_5l}).
Hence we can omit these four functions and derive relations
between the polarizabilities by comparing the amplitudes $A_{i}$
of the two different low-energy expansions (\ref{gl3_4a})--(\ref{gl3_4l})
and (\ref{gl3_5a})--(\ref{gl3_5l}) of
${\mathcal{M}}_{B}^{\gamma^{\ast} \gamma}$.
To be specific, $A_{4}$ in Eq. (\ref{gl3_5d}) vanishes to lowest order in
$\omega'$, thus relating the polarizabilities
$P^{(11,11)1}$, $P^{(11,02)1}$,  and $\hat{P}^{(11,2)1}$.
Two further relations arise because the terms with and without 
$\cos \theta$ in the amplitudes $A_{1}$ and $A_{3}$ are, respectively, 
given by the same linear combinations of the $f_{i}$.
An inspection of $A_{11}$ and $A_{12}$ yields a fourth
relation:
While Eqs. (\ref{gl3_5k}) and (\ref{gl3_5l}) contain only two
independent linear combinations of the $f_{i}$, $A_{11}$ and $A_{12}$
in Eqs. (\ref{gl3_4k}) and (\ref{gl3_4l}) depend on three
polarizabilities.
Note that the identity $\omega_{0}^{2} = \bar{q}^{2} + 2 M \omega_{0}$
enters into the derivation of the last relation.
Altogether, we obtain four relations between the ten original generalized 
polarizabilities,
\begin{mathletters}
\label{gl3_6}
\begin{eqnarray} \label{gl3_6a}
0 & = & \sqrt{\frac{3}{2}} P^{(01,01)0}(\bar{q}^2)
 + \sqrt{\frac{3}{8}} P^{(11,11)0}(\bar{q}^2)
 + \frac{3 \bar{q}^2}{2 \omega_0} \hat{P}^{(01,1)0}(\bar{q}^2) \, ,
\\ \label{gl3_6b}
0 & = & P^{(11,11)1}(\bar{q}^2)
 + \sqrt{\frac{3}{2}} \omega_0 P^{(11,02)1}(\bar{q}^2)
 + \sqrt{\frac{5}{2}} \bar{q}^2 \hat{P}^{(11,2)1}(\bar{q}^2) \, ,
\\ \label{gl3_6c}
0 & = & 2 \omega_0 P^{(01,01)1}(\bar{q}^2)
 + 2 \frac{\bar{q}^2}{\omega_0} P^{(11,11)1}(\bar{q}^2)
 - \sqrt{2} \bar{q}^2 P^{(01,12)1}(\bar{q}^2)
 + \sqrt{6} \bar{q}^2 \hat{P}^{(01,1)1}(\bar{q}^2) \, ,
\\ \label{gl3_6d}
0 & = & 3 \frac{\bar{q}^2}{\omega_0} P^{(01,01)1}(\bar{q}^2)
 - \sqrt{3} P^{(11,00)1}(\bar{q}^2)
 - \sqrt{\frac{3}{2}} \bar{q}^2 P^{(11,02)1}(\bar{q}^2) \, .
\end{eqnarray}
\end{mathletters}
As is evident from the definition of the generalized polarizabilities
in Eqs.\ (\ref{gl3_1a}) and (\ref{gl3_1b}) the relations 
(\ref{gl3_6a})--(\ref{gl3_6d}) can only be applied along the VCS line
$\omega' = 0$.
The relation between the scalar polarizabilities in Eq. (\ref{gl3_6a})
has already been derived in Ref.\ \cite{Drechsel_96}.
It is an important consequence of Eqs.\ (\ref{gl3_6a})--(\ref{gl3_6d})
that six independent functions of $\bar{q}^2$ are sufficient to 
parameterize the structure-dependent VCS amplitude to lowest order 
in $\omega'$.

We want to emphasize again that the four relations between the ten 
generalized polarizabilities are ultimately caused by charge conjugation 
in connection with nucleon crossing.
If we drop the assumption that this symmetry holds, the functions
$f_{3}$, $f_{4}$, $f_{8}$, and $f_{10}$ give a contribution to the
leading-order terms in Eqs.\ (\ref{gl3_5a})--(\ref{gl3_5l}), and none
of our four relations between the polarizabilities is valid any longer.
In this sense we find complete agreement with the analysis of Ref.\  
\cite{Guichon_95}, because the constraint due to charge conjugation 
and nucleon crossing has not been taken care of in that reference.

At $\bar{q} = 0$, particular relations between the polarizabilities
and their derivatives can be found by expanding Eqs.
(\ref{gl3_6a})--(\ref{gl3_6d}).
We only discuss the most interesting cases:
Three of the seven vector polarizabilities vanish
at $\bar{q} = 0$,
\begin{equation} \label{gl3_7}
P^{(01,01)1}(0) = P^{(11,11)1}(0) = P^{(11,00)1}(0) = 0 \, .
\end{equation}
These results follow, in part, from Eqs.\ (\ref{gl3_6b}) and (\ref{gl3_6d}),
if one exploits the expansion
$\omega_{0} = - \bar{q}^{2}/2M + {\mathcal{O}}(\bar{q}^{4})$.
Equation (\ref{gl3_6d}) only contains the information that a certain linear 
combination of $P^{(01,01)1}(0)$ and $P^{(11,00)1}(0)$ disappears.
The fact that both polarizabilities vanish separately becomes obvious
by comparing the angular-independent part of the amplitude 
$A_{11}$ in Eqs.\ (\ref{gl3_4k}) and (\ref{gl3_5k}).

Combining Eqs.\ (\ref{gl3_6b}) and (\ref{gl3_6c}) enables us to
eliminate $P^{(11,11)1}$.
This leads to a relation between the remainig four vector polarizabilities,
\begin{equation} \label{gl3_8}
P^{(01,12)1}(0) + \sqrt{3} P^{(11,02)1}(0)
- \sqrt{3} \hat{P}^{(01,1)1}(0) - 2 \sqrt{5} M \hat{P}^{(11,2)1}(0) =
0 \, .
\end{equation}

The relations between the generalized polarizabilities also imply
that several multipoles are connected at small values of $\omega'$.
Making use of Eqs. (\ref{gl3_6a}) and (\ref{gl3_6b}) we list the two
most striking examples,
\begin{mathletters}
\begin{eqnarray} \label{gl3_9a}
H^{(21,21)0}(\omega' , \bar{q}) & = &
\omega' \Bigl [ 2 \omega_{0} P^{(01,01)0}(\bar{q}^2)
+ \sqrt{6} \bar{q}^{2} \hat{P}^{(01,1)0}(\bar{q}^2) \Bigr ]
+ {\mathcal{O}}(\omega'^{2})
\nonumber\\
& = & - \omega' \omega_{0} P^{(11,11)0}(\bar{q}^2)
+ {\mathcal{O}}(\omega'^{2}) \vphantom{\frac{1}{1}}
\nonumber\\
& = & - \frac{\omega_{0}}{\bar{q}} H^{(11,11)0}(\omega' , \bar{q})
+ {\mathcal{O}}(\omega'^{2}) \, ,
\\ \label{gl3_9b}
H^{(11,11)1}(\omega' , \bar{q}) & = &
\omega' \bar{q} P^{(11,11)1}(\bar{q}^2)
+ {\mathcal{O}}(\omega'^{2})  \vphantom{\frac{1}{1}}
\nonumber\\
& = & \omega' \biggl [
- \sqrt{\frac{3}{2}} \omega_{0} \bar{q} P^{(11,02)1}(\bar{q}^2)
- \sqrt{\frac{5}{2}} \bar{q}^{3} \hat{P}^{(11,2)1}(\bar{q}^2) \biggr]
+ {\mathcal{O}}(\omega'^{2})
\nonumber\\
& = & H^{(11,22)1}(\omega' , \bar{q})
+ {\mathcal{O}}(\omega'^{2})  \vphantom{\frac{1}{1}} \, .
\end{eqnarray}
\end{mathletters}
These equations are based upon the low-energy expansion of the multipoles 
given in Ref.\ \cite{Guichon_95}.
Obviously, charge conjugation leads, at least in VCS, to unexpected
constraints between the multipoles, which go beyond the conditions due
to parity and angular momentum conservation.
Whether these constraints are limited to the lowest order in $\omega'$
is beyond the scope of our present investigation.
An answer to this question would require both a multipole analysis including
angular momenta $L' \geq 2$, and an extension of 
Eqs. (\ref{gl3_5a})--(\ref{gl3_5l}) to higher orders in $\omega'$.

In Ref.\ \cite{Drechsel_96} it has been argued that the relation between the 
scalar electric and magnetic multipole (Eq. (\ref{gl3_9a})) vanishes in 
the static limit $M \to \infty$, which is obvious from the definition of 
$\omega_0$.
However, the second equation (\ref{gl3_9b}) is not affected by this
limit.
Accordingly, while Eq. (\ref{gl3_9a}) may be interpreted as a recoil effect,
the connection between $H^{(11,11)1}$ and $H^{(11,22)1}$ seems to indicate
an intrinsic property of the target.

   From a practical point of view, the results in Eqs.
(\ref{gl3_6a})--(\ref{gl3_6d}) are very appropriate to test predictions 
for the generalized polarizabilities of models incorporating the required
symmetries.
Moreover, they can serve as constraints for experimental analyses.

With the exception of the electric polarizability $\alpha(\bar{q}^2)$,
the measurement of individual polarizabilities requires polarization
experiments.
In the unpolarized case it has been proposed \cite{Guichon_95}
to extract four linear combinations of the polarizabilities by
measuring the structure functions
\begin{mathletters}
\begin{eqnarray} \label{gl3_10a}
P_{LL}(\bar{q}) & = &
-2 \sqrt{6} M G_{E}(Q_{0}^{2}) P^{(01,01)0}(\bar{q}^2) \, ,
\vphantom{\frac{1}{1}}
\\ \label{gl3_10b}
P_{TT}(\bar{q}) & = &
\frac{3}{2} G_{M}(Q_{0}^{2})
\Bigl [ 2 \omega_{0} P^{(01,01)1}(\bar{q}^2)
+ \sqrt{2} \bar{q}^{2} \Bigl ( P^{(01,12)1}(\bar{q}^2)
              + \sqrt{3} \hat{P}^{(01,1)1}(\bar{q}^2) \Bigr ) \Bigr ] \, ,
\\ \label{gl3_10c}
P_{LT}(\bar{q}) & = &
\sqrt{\frac{3}{2}} \frac{M \bar{q}}{\sqrt{Q_{0}^{2}}}
G_{E}(Q_{0}^{2}) P^{(11,11)0}(\bar{q}^2)
\nonumber \\
& & 
+ \frac{\sqrt{3} \sqrt{Q_{0}^{2}}}{2 \bar{q}} G_{M}(Q_{0}^{2})
\Bigl ( P^{(11,00)1}(\bar{q}^2)
+ \frac{\bar{q}^{2}}{\sqrt{2}} P^{(11,02)1}(\bar{q}^2) \Bigr ) \, ,
\\ \label{gl3_10d}
P_{LT}'(\bar{q}) & = &
\sqrt{\frac{3}{2}} \frac{M}{\sqrt{Q_{0}^{2}}}
G_{E}(Q_{0}^{2}) \Bigl ( 2 \omega_{0} P^{(01,01)0}(\bar{q}^2)
+ \sqrt{6} \bar{q}^{2} \hat{P}^{(01,1)0}(\bar{q}^2) \Bigr)
\nonumber \\
& & 
- \frac{3}{2} \sqrt{Q_{0}^{2}} G_{M}(Q_{0}^{2}) P^{(01,01)1}(\bar{q}^2) \, ,
\end{eqnarray}
\end{mathletters}
with $G_{E}$ and $G_{M}$ denoting the electric and magnetic Sachs form 
factors, respectively.
These structure functions describe, to lowest order in $\omega'$, the 
interference between the non-Born and the Born plus Bethe-Heitler 
amplitude.
By use of Eqs. (\ref{gl3_6a}) and (\ref{gl3_6d}) the
structure functions $P_{LT}$ and $P_{LT}'$ turn out to be
mutually dependent via the relation
\begin{equation} \label{gl3_11}
P_{LT}(\bar{q}) + \frac{\bar{q}}{\omega_{0}} P_{LT}'(\bar{q}) = 0 \, .
\end{equation}
This indicates that in an unpolarized experiment there are only three
independent structure functions containing five generalized polarizabilities.

\section{Summary}
\label{chapter_4}
We analyzed VCS off the nucleon in a covariant, model-independent
formalism, which allowed us to include constraints from discrete 
symmetries in a natural way. 
We restricted our investigation to the so-called structure-dependent part 
which is obtained from the full amplitude by subtracting a separately 
gauge-invariant Born part involving the vertex of Eq.\ (\ref{f1f2vertex}).
We demonstrated that it is possible to parametrize the VCS invariant 
matrix element in such a fashion that the tensor structures as well 
as the corresponding amplitudes are free of kinematical singularities. 
Consequently, the amplitudes only contain information on the dynamics 
of the process to be explored by the experiment. 
We then focused on Compton scattering with a virtual, spacelike photon 
in the initial and a real photon in the final state, because this process 
will be investigated in future experiments.
Applying our covariant approach to particular kinematical scenarios we
critically reviewed the formalism presently used in the analysis of
VCS experiments below pion threshold \cite{Guichon_95}. 
We found that charge-conjugation symmetry in connection with nucleon
crossing generates four relations among the ten originally proposed 
generalized polarizabilities of the nucleon. 
We further derived relations between the generalized polarizabilities at
particular kinematical points. 
We hope that our results will facilitate future theoretical and 
experimental analysis.
These results have already been quite valuable for the analysis of VCS
within the framework of the linear sigma model \cite{Metz_96,Metz_96a}
and HBChPT \cite{Hemmert_96a,Hemmert_96c}.
All constraints on the generalized polarizabilities derived in this paper 
were confirmed on the level of model calculations with these two effective 
Lagrangians, because they incorporate the relevant symmetries, gauge 
invariance and Lorentz invariance as well as the discrete symmetries.
We consider this as an important check for both the model calculations 
and our general results.

\acknowledgments
A.\ Yu.\ K.\ would like to thank the theory group of the Institut f\"{u}r
Kernphysik for the hospitality during his visit in Mainz.

\appendix
\section{General form of the Compton tensor}
\label{appendix_a}
A construction of the Compton tensor $M^{\mu\nu}$ of the most general 
VCS reaction $\gamma^{\ast} + N \to \gamma^{\ast} + N$ has been given 
by Tarrach \cite{Tarrach_75}.
Here, we sketch the main features of his derivation and extend it with
respect to our considerations.
The list of all possible tensor structures $K^{\mu\nu}_i$ of the most
general Compton tensor is built up from the four independent Lorentz 
vectors $q^{\mu}$, $q'^{\mu}$, $P^{\mu}$, $\gamma^{\mu}$.
Each structure $K^{\mu\nu}_i$ must be even with respect to parity 
transformations, because we consider only parity-conserving interactions.
Furthermore, it is useful to choose the $K_{i}^{\mu\nu}$ with a 
well-defined behavior under photon crossing
($q \leftrightarrow - q' , \; \mu \leftrightarrow \nu$) and under the 
combination of nucleon crossing and charge conjugation $C$.
With these assumptions one obtains 34 $K^{\mu\nu}_i$
(see Eq.\ (8) of \cite{Tarrach_75} for the complete list)
\begin{equation} \label{gla_1}
K_1^{\mu\nu}=g^{\mu\nu} \, , \quad \cdots,\quad
K_{34}^{\mu\nu} = 
(\gamma^{\mu} \gamma^{\nu} - \gamma^{\nu} \gamma^{\mu}) Q \cdot \gamma
+ Q \cdot \gamma
(\gamma^{\mu} \gamma^{\nu} - \gamma^{\nu} \gamma^{\mu})
\,, 
\end{equation}
   where the structures $K^{\mu\nu}_1$ - $K^{\mu\nu}_{10}$ would also
appear in the derivation for a spin-0 particle.
   Note our definition of $P$ and $Q$ of Eq.\ (\ref{gl2_1}) and the
reversed order of $\mu$ and $\nu$ as compared with Ref.\ \cite{Tarrach_75}.
   Using four-momentum conservation and Dirac's equation it is possible 
to express each other tensor in terms of the $K_{i}^{\mu\nu}$.
Moreover, two nontrivial relations between several of the
$K_{i}^{\mu\nu}$ hold \cite{Tarrach_75}, reducing the number of
independent tensors to 32.
Even if there is some freedom in the choice of the independent tensors,
it is convenient to eliminate $K_{13}^{\mu\nu}$ and
$K_{28}^{\mu\nu}$ \cite{Tarrach_75}, which will not appear in the
following derivation any more.
Counting the helicities of the four particles involved in the reaction
one ends up with the same number
$32 = (2 \times 2 \times 4 \times 4)/2$, where the division by 2 is due
to parity conservation in boson-fermion scattering \cite{Fearing_84}.
Since each photon is considered off shell, it has components
of spin 1 and spin 0 and thus enters with four degrees of freedom
into the counting \cite{Fearing_84}.

In order to incorporate current conservation at both photon vertices
[see Eq. (\ref{gl2_4})], one derives linear combinations of the 
$K_{i}^{\mu\nu}$, which then form the basis vectors of $M^{\mu\nu}$.
For the spin-independent amplitude this procedure has been explained
in more detail in Refs.\ \cite{Fearing_96} and \cite{Drechsel_96}.
In the construction of such gauge-invariant linear combinations it 
usually happens that poles in the independent invariants
$q^{2}$, $q'^{2}$, $q \cdot q'$, and $q \cdot P$
of the VCS reaction arise, leading to unphysical zeros or constraints 
in the corresponding amplitudes of the basis vectors.
A general solution developed by Bardeen and Tung \cite{Bardeen_68} avoids
this problem, which one encounters in different physical reactions.
The application of this method to VCS results in 18 gauge-invariant and 
pole-free tensors \cite{Tarrach_75},
\begin{mathletters}
\label{gla_2}
\begin{eqnarray} \label{gla_2a}
T_1^{\mu\nu}
& = &
- q \cdot q' K_1^{\mu\nu} + K_3^{\mu\nu}
\, ,\cdots, \\ 
\label{gla_2r}
T_{18}^{\mu\nu}
& = &
K_{17}^{\mu\nu} - 2 q \cdot P K_{25}^{\mu\nu}
+ \frac{q \cdot q'}{2} K_{34}^{\mu\nu}
\, .
\end{eqnarray}
\end{mathletters}
The spin-independent tensors $T_{1}^{\mu\nu}, \ldots , T_{5}^{\mu\nu}$
are the same as in Eq.\ (5) of Ref.\ \cite{Drechsel_96}, 
whereas the basis elements
$T_{6}^{\mu\nu}, \ldots , T_{18}^{\mu\nu}$ correspond to the tensors
$\tau_{6}^{\mu\nu}, \ldots , \tau_{18}^{\mu\nu}$ in Eq.\ (12)
of Ref.\ \cite{Tarrach_75}, rewritten for our choice of $P$ and $Q$ in 
Eq.\ (\ref{gl2_1}). 
Note that the number of these tensors also results from counting
one longitudinal and two transverse degrees of polarization of the
virtual photons, $18 = (2 \times 2 \times 3 \times 3)/2$.

The above considerations determine the general form of $M^{\mu\nu}$.
In particular, the gauge invariant residual part $M_{B}^{\mu\nu}$
[see Eq.\ (\ref{gl2_3})] of the Compton tensor can be expressed
in terms of the basis vectors in (\ref{gla_2a}) and (\ref{gla_2r})
according to
\begin{equation} \label{gla_3}
M_{B}^{\mu\nu} = \sum_{i=1}^{18}
B_{i}' (q^{2} , q'^{2} , q \cdot q' , q \cdot P) T_{i}^{\mu\nu} \, .
\end{equation}

However, the above basis has one drawback.
Though the tensors $T_{i}^{\mu\nu}$ are free of poles, the corresponding
amplitudes $B_{i}'$ still contain kinematical constraints.
Such a basis is called ``nonminimal'' \cite{Tarrach_75}.
The nonminimality is due to the fact that it is impossible to
make a transformation into an equivalent, pole-free basis without
introducing any kinematical pole in the transformation matrix
\cite{Tarrach_75}.
As a consequence, three further gauge-invariant and pole-free tensors 
exist, which can be obtained from
$T_{1}^{\mu\nu}, \ldots , T_{18}^{\mu\nu}$ only with factors carrying
a single pole in $q \cdot q'$:
\begin{mathletters}
\begin{eqnarray} \label{gla_4a}
T_{19}^{\mu\nu} & = &
\frac{1}{q \cdot q'} \left[ - q^2 q'^2 T_2^{\mu\nu}
+ (q \cdot P)^2 T_3^{\mu\nu} - q \cdot P \frac{q^2 + q'^2}{2} T_4^{\mu\nu}
+ q \cdot P \frac{q^2 - q'^2}{2} T_5^{\mu\nu} \right]
\nonumber\\
& = &
(q \cdot P)^2 K_2^{\mu\nu} + q^2 q'^2 K_6^{\mu\nu}
- q \cdot P \frac{q^2 + q'^2}{2} K_9^{\mu\nu}
- q \cdot P \frac{q^2 - q'^2}{2} K_{10}^{\mu\nu}
\, , \\ \label{gla_4b}
T_{20}^{\mu\nu} & = &
\frac{1}{4 q \cdot q'} \left[ (q^2 - q'^2) T_{10}^{\mu\nu}
- 2 (q^2 + q'^2) T_{14}^{\mu\nu} + 2 q \cdot P T_{15}^{\mu\nu} \right]
\nonumber\\
& = &
- \frac{q^2 - q'^2}{2} K_6^{\mu\nu} - \frac{q \cdot P}{2} K_{10}^{\mu\nu}
+ M \frac{q^2 - q'^2}{2} K_{21}^{\mu\nu}
+ M \frac{q^2 + q'^2}{2} K_{22}^{\mu\nu}
- M q \cdot P K_{24}^{\mu\nu}
\nonumber \\
& &
+ \frac{q^2 + q'^2}{8}  K_{27}^{\mu\nu} - \frac{q \cdot P}{4} K_{29}^{\mu\nu}
- q \cdot P \frac{q^2 - q'^2}{4} K_{33}^{\mu\nu}
+ M \frac{q^2 - q'^2}{8} K_{34}^{\mu\nu}
\, , \\ \label{gla_4c}
T_{21}^{\mu\nu} & = &
\frac{1}{4 q \cdot q'} \left[ (q^2 + q'^2) T_{10}^{\mu\nu}
- 2 (q^2 - q'^2) T_{14}^{\mu\nu} + 2 q \cdot P T_{16}^{\mu\nu} \right]
\nonumber\\
& = &
- \frac{q^2 + q'^2}{2} K_6^{\mu\nu} + \frac{q \cdot P}{2} K_9^{\mu\nu}
+ M \frac{q^2 + q'^2}{2} K_{21}^{\mu\nu}
+ M \frac{q^2 - q'^2}{2} K_{22}^{\mu\nu}
- M q \cdot P K_{23}^{\mu\nu}
\nonumber \\
& &
+ \frac{q^2 - q'^2}{8}  K_{27}^{\mu\nu} - \frac{q \cdot P}{4} K_{30}^{\mu\nu}
- q \cdot P \frac{q^2 + q'^2}{4} K_{33}^{\mu\nu}
+ M \frac{q^2 + q'^2}{8} K_{34}^{\mu\nu}
\, .
\end{eqnarray}
\end{mathletters}
The nonminimality of the basis in Eq.\ (\ref{gla_2}) is reflected by
the fact that in the case $q \cdot q' = 0 $ the set of tensors in 
Eq.\ (\ref{gla_2}) does not form a tensor basis any more, because some 
elements of the original basis become linearly dependent \cite{Tarrach_75}.
Unfortunately, the two kinematical scenarios we investigate for the 
analysis of VCS at small final photon energy $\omega'$ both imply 
$q \cdot q' = 0$.

   For this reason, when constructing the tensor basis for the residual part 
$M_B^{\mu\nu}$, we will have to start with a tensor basis different from the 
one of Eq.\ (\ref{gla_2}).

It turns out that if we use $T_{19}^{\mu\nu}$ instead of 
$T_5^{\mu\nu}$, $T_{20}^{\mu\nu}$ instead of $T_{15}^{\mu\nu}$, and 
$T_{21}^{\mu\nu}$ instead of $T_{16}^{\mu\nu}$, we obtain a tensor basis 
which is free of poles and zeroes and, thus, can also be used in the 
case $q \cdot q' = 0$. 
However, this new basis is not minimal either, because poles in the 
invariant $q \cdot P$ can create linear dependences among the basis 
elements in the Born part of the Compton tensor. 
However, this is not the case for the residual part, which we are 
interested in in this paper. 
The residual part of the Compton tensor reads
\begin{equation} \label{gla_neu}
M_{B}^{\mu \nu} = \sum_{i \in J}
 B_i (q^2 , q'^2 , q\cdot q' , q \cdot P) T_i^{\mu \nu} \,, \,\,\,
J = \left\{1, \ldots , 21 \right\} \backslash \left\{ 5, 15, 16\right\} \, .
\end{equation}
The corresponding amplitudes $ B_i (q^2 , q'^2 , q\cdot q' , q \cdot
P)$ are free of kinematical constraints, in particular free of
poles. 
This can be proved by means of considering their symmetry properties:
The tensor $M_{B}^{\mu\nu}$ is invariant under photon crossing and the 
combination of charge conjugation with nucleon crossing \cite{Tarrach_75}.
Since the $T_i^{\mu\nu}$ exhibit definite transformation properties
with respect to photon crossing and charge conjugation combined with
nucleon crossing, the amplitudes $B_{i}$ do as well.
By means of the identities
\begin{mathletters}
\label{gl2_6}
\begin{eqnarray} \label{gl2_6a}
B_{i}(q^{2} , q'^{2} , q \cdot q' , q \cdot P) & = &
+ B_{i}(q'^{2} , q^{2} , q \cdot q' , - q \cdot P)
\nonumber \\
& & 
\quad (i = 1,2,3,5,8,10,13,15,18,19,21) \, ,
\\ \label{gl2_6b}
B_{i}(q^{2} , q'^{2} , q \cdot q' , q \cdot P) & = &
- B_{i}(q'^{2} , q^{2} , q \cdot q' , - q \cdot P)
\nonumber \\
& & 
\quad (i = 4,6,7,9,11,12,14,16,17,20) \, ,
\\ \label{gl2_6c}
B_{i}(q^{2} , q'^{2} , q \cdot q' , q \cdot P) & = &
+ B_{i}(q^{2} , q'^{2} , q \cdot q' , - q \cdot P)
\nonumber \\
& & 
\quad (i = 1,2,3,8,9,10,11,14,18,19,20,21) \, ,
\\ \label{gl2_6d}
B_{i}(q^{2} , q'^{2} , q \cdot q' , q \cdot P) & = &
- B_{i}(q^{2} , q'^{2} , q \cdot q' , - q \cdot P)
\nonumber \\
& & 
\quad (i = 4,5,6,7,12,13,15,16,17) \, ,
\end{eqnarray}
\end{mathletters}
the functions $B_{i}$ can be divided into four classes, where
in Eqs.\ (\ref{gl2_6a}) and (\ref{gl2_6b}) use has been made of the identity
$q \cdot P = q' \cdot P$.
We emphasize that Eqs.\ (\ref{gl2_6c}) and (\ref{gl2_6d}), which are crucial 
for the derivation of the relations between the generalized
polarizabilities in Sec. \ref{chapter_3}, may alternatively be derived 
by means of time reversal together with photon crossing \cite{Tarrach_75}.

For the definition of low-energy constants we
need a general expansion of the $B_{i}$ up to the order
${\mathcal{O}}(k^{3})$ ($k \in \{q,q' \}$), which immediately follows from 
the transformation properties of Eqs. (\ref{gl2_6a})--(\ref{gl2_6d}):
\begin{mathletters}
\label{gl2_7}
\begin{eqnarray} \label{gl2_7a}
B_i & = & b_{i,0} + b_{i,2a} q \cdot q'
+  b_{i,2b} (q^2 + q'^2) + b_{i,2c} (q \cdot P)^2
+ {\mathcal{O}}(k^{4})
\nonumber \\
& & 
\quad (i=1,2,3,8,10,18,19,21) \, ,
\\ \label{gl2_7b}
B_i & = & b_{i,3} (q^2 - q'^2) q \cdot P
+ {\mathcal{O}}(k^{4})
\quad (i=5,13,15) \, ,
\\ \label{gl2_7c}
B_i & = & b_{i,2} (q^2 - q'^2)
+ {\mathcal{O}}(k^{4})
\quad (i=9,11,14,20) \, ,
\\ \label{gl12_7d}
B_i & = & b_{i,1} q \cdot P + b_{i,3a} q \cdot P q \cdot q' +
b_{i,3b} q \cdot P (q^2 + q'^2) + b_{i,3c} (q \cdot P)^3
+ {\mathcal{O}}(k^{4})
\nonumber \\
& & 
\quad (i=4,6,7,12,16,17)
\, .
\end{eqnarray}
\end{mathletters}
Such an expansion of the amplitudes in terms of the four-momenta
of the photons has already been performed in Ref.\ \cite{Fearing_96}
in connection with VCS from the pion.

   From the above Taylor expansion, the fact that in the original 
representation
\begin{equation}
M_B^{\mu\nu} = \sum_{r \in R} C_r (q^2,q'^2,q \cdot q',q \cdot P)
K_r^{\mu\nu}\, , \,\,\, R = \left\{1, \ldots, 34 \right\} \backslash \left\{
13, 28 \right\},
\end{equation}
the functions $C_r$ by definition are free of poles in the kinematical 
variables, and the symmetry properties of the $C_r$ and $K_r^{\mu\nu}$
it follows that the functions
$ B_i (q^2 , q'^2 , q\cdot q' , q \cdot P) \, , \,\,\, i \in J ,$ are
free of poles. 
Furthermore, it can be shown that gauge invariance does not generate any 
additional kinematical constraints on these functions. 
Thus, Eq. (\ref{gla_neu}) contains a representation for $M_B^{\mu\nu}$ 
which satisfies all requirements --- not only for our particular case 
$q \cdot q' = 0$, but for any choice of kinematical variables in 
$\gamma^* N \rightarrow \gamma^* N$. 
In particular, it is not necessary to use 3 additional functions as in 
\cite{Tarrach_75}.
Reexpressing this parametrization in the form of Eq. (\ref{gla_3}), 
the functions $B'_i$ read
\begin{mathletters}
\begin{eqnarray} \label{gla_7a}
B_i' & = & B_i \,\,\,\, \mathrm{for} \,\,\,\, i \in \{
1,6,7,8,9,11,12,13,17,18\} \, ,
\\ \label{gla_7b}
B_2' & = & B_2 - \frac{q^2 q'^2}{q \cdot q'} B_{19} \, ,
\\ \label{gla_7c}
B_3' & = & B_3 + \frac{(q \cdot P)^2}{q \cdot q'} B_{19} \, ,
\\ \label{gla_7d}
B_4' & = & B_4 - q \cdot P \frac{q^2 + q'^2}{2 q \cdot q'} B_{19} \, ,
\\ \label{gla_7e}
B_5' & = & q \cdot P \frac{q^2 - q'^2}{2 q \cdot q'} B_{19} \, ,
\\ \label{gla_7f}
B_{10}' & = & B_{10} + \frac{q^2 - q'^2}{4 q \cdot q'} B_{20}
+ \frac{q^2 + q'^2}{4 q \cdot q'} B_{21} \, ,
\\ \label{gla_7g}
B_{14}' & = & B_{14} - \frac{q^2 + q'^2}{2 q \cdot q'} B_{20}
- \frac{q^2 - q'^2}{2 q \cdot q'} B_{21}\, ,
\\ \label{gla_7h}
B_{15}' & = & \frac{q \cdot P}{2 q \cdot q'} B_{20} \, ,
\\ \label{gla_7i}
B_{16}' & = & \frac{q \cdot P}{2 q \cdot q'} B_{21} \, .
\end{eqnarray}
\end{mathletters}
These equations follow from the definitions of
$T_{19}^{\mu\nu}$, $T_{20}^{\mu\nu}$, and $T_{21}^{\mu\nu}$ in
Eqs.\ (\ref{gla_4a})--(\ref{gla_4c}). \footnote{Note
that $B_{19}$ is equivalent to the function $B_{6}$ in Ref.\ 
\cite{Drechsel_96}.}
We stress that the tensors $T_{1}^{\mu\nu}, \ldots , T_{18}^{\mu\nu}$
still form a basis of the Compton tensor according to Eq. (\ref{gla_3}).
The nonminimality of this basis is expressed in a specific kinematical
behavior of the amplitudes $B_{i}'$, namely, some amplitudes contain
poles in $q \cdot q'$.
However, $M_{B}^{\mu\nu}$ is free of poles, despite the behavior of
the $B_{i}'$.
This is due to the fact that both the amplitudes $B_{i}$ and the tensors
$T_{i}^{\mu\nu}\,, \,\,\, i \in J ,$ do not carry any pole in
the relativistic invariants.

For the discussion of ${\cal M}^{\gamma^*\gamma}$ we change the numbering 
by introducing tensors $\rho_{i}^{\mu\nu}$ in the following way:
\begin{mathletters}
\label{gl2_8}
\begin{eqnarray} \label{gl2_8a}
  \varepsilon_{\mu} \rho_1^{\mu \nu} \varepsilon'^*_{\nu}
& = & \varepsilon_{\mu} \tilde{T}_1^{\mu \nu} \varepsilon'^*_{\nu}
\nonumber\\
& = & \varepsilon \cdot q' \varepsilon'^* \cdot q
- q \cdot q' \varepsilon \cdot \varepsilon'^* \, ,
\\ \label{gl2_8b}
  \varepsilon_{\mu} \rho_2^{\mu \nu} \varepsilon'^*_{\nu}
& = & \varepsilon_{\mu} \tilde{T}_2^{\mu \nu} \varepsilon'^*_{\nu}
\nonumber\\
& = & q\cdot P ( \varepsilon \cdot P \varepsilon'^* \cdot q
+ \varepsilon'^* \cdot P \varepsilon \cdot q' )
- q \cdot q' \varepsilon \cdot P \varepsilon'^* \cdot P
- (q \cdot P)^2 \varepsilon \cdot \varepsilon'^* \, ,
\\ \label{gl2_8c}
  \varepsilon_{\mu} \rho_3^{\mu \nu} \varepsilon'^*_{\nu}
& = & \varepsilon_{\mu} \tilde{T}_4^{\mu \nu} \varepsilon'^*_{\nu}
\nonumber\\
& = & q \cdot P q^2 \varepsilon \cdot \varepsilon'^*
- q \cdot P \varepsilon \cdot q \varepsilon'^* \cdot q
- q^2 \varepsilon'^* \cdot P \varepsilon \cdot q'
+ q \cdot q' \varepsilon'^* \cdot P \varepsilon \cdot q \, ,
\\ \label{gl2_8d}
  \varepsilon_{\mu} \rho_4^{\mu \nu} \varepsilon'^*_{\nu}
& = & \varepsilon_{\mu} \tilde{T}_7^{\mu \nu} \varepsilon'^*_{\nu}
\nonumber\\
& = & \varepsilon \cdot P \varepsilon'^* \cdot P Q \cdot \gamma
- q \cdot P (\varepsilon \cdot P \varepsilon'^* \cdot \gamma
+ \varepsilon'^* \cdot P \varepsilon \cdot \gamma)
\nonumber \\
& & 
+ i q \cdot P \gamma_5 \varepsilon^{\mu\nu\alpha\beta}
\varepsilon_{\mu} \varepsilon'^*_{\nu} Q_{\alpha} \gamma_{\beta} \, ,
\\ \label{gl2_8e}
  \varepsilon_{\mu} \rho_5^{\mu \nu} \varepsilon'^*_{\nu}
& = & \varepsilon_{\mu} \tilde{T}_8^{\mu \nu} \varepsilon'^*_{\nu}
= - \varepsilon_{\mu} \tilde{T}_9^{\mu \nu} \varepsilon'^*_{\nu}
\nonumber\\
& = & \frac{1}{4} \varepsilon'^* \cdot P \varepsilon \cdot q Q \cdot \gamma
+ \frac{q^2}{4} ( \varepsilon \cdot P \varepsilon'^* \cdot \gamma
- \varepsilon'^* \cdot P \varepsilon \cdot \gamma)
- \frac{q \cdot P}{2} \varepsilon \cdot q \varepsilon'^* \cdot \gamma
\nonumber \\
& & 
+ \frac{i}{4} q^2 \gamma_5 \varepsilon^{\mu\nu\alpha\beta}
\varepsilon_{\mu} \varepsilon'^*_{\nu} Q_{\alpha} \gamma_{\beta} \, ,
\\ \label{gl2_8f}
  \varepsilon_{\mu} \rho_{6}^{\mu \nu} \varepsilon'^*_{\nu}
& = & \varepsilon_{\mu} \tilde{T}_{10}^{\mu \nu} \varepsilon'^*_{\nu}
\nonumber\\
& = & - 2 q \cdot q' \varepsilon \cdot P \varepsilon'^* \cdot P
+ q \cdot P ( \varepsilon \cdot P \varepsilon'^* \cdot q
+ \varepsilon'^* \cdot P \varepsilon \cdot q')
\nonumber \\
& & 
+ 2 M q \cdot q' (\varepsilon \cdot P \varepsilon'^* \cdot \gamma
+ \varepsilon'^* \cdot P \varepsilon \cdot \gamma)
- 2 M q \cdot P (\varepsilon \cdot q' \varepsilon'^* \cdot \gamma
+ \varepsilon'^* \cdot q \varepsilon \cdot \gamma)
\nonumber \\
& & 
+ i q \cdot P (\varepsilon \cdot q' \sigma^{\nu \alpha}
\varepsilon'^*_{\nu} Q_{\alpha}
- \varepsilon'^* \cdot q \sigma^{\mu \alpha} \varepsilon_{\mu} Q_{\alpha} )
+ 2 i q \cdot q' q \cdot P \sigma^{\mu \nu}
\varepsilon_{\mu} \varepsilon'^*_{\nu}
\nonumber \\
& & 
+ 2 i M q \cdot q' \gamma_5 \varepsilon^{\mu \nu \alpha \beta}
\varepsilon_{\mu} \varepsilon'^*_{\nu} Q_{\alpha} \gamma_{\beta} \, ,
\\ \label{gl2_8g}
  \varepsilon_{\mu} \rho_{7}^{\mu \nu} \varepsilon'^*_{\nu}
& = & \varepsilon_{\mu} \tilde{T}_{11}^{\mu \nu} \varepsilon'^*_{\nu}
\nonumber\\
& =& \frac{1}{4} (\varepsilon \cdot P \varepsilon'^* \cdot q
- \varepsilon'^* \cdot P \varepsilon \cdot q' ) Q \cdot \gamma
- \frac{q \cdot q'}{2} (\varepsilon \cdot P \varepsilon'^* \cdot \gamma
- \varepsilon'^* \cdot P \varepsilon \cdot \gamma)
\nonumber \\
& & 
+ \frac{q \cdot P}{2} (\varepsilon \cdot q' \varepsilon'^* \cdot \gamma
- \varepsilon'^* \cdot q \varepsilon \cdot \gamma) \, ,
\\ \label{gl2_8h}
  \varepsilon_{\mu} \rho_{8}^{\mu \nu} \varepsilon'^*_{\nu}
& = & \varepsilon_{\mu} \tilde{T}_{12}^{\mu \nu} \varepsilon'^*_{\nu}
= \varepsilon_{\mu} \tilde{T}_{13}^{\mu \nu} \varepsilon'^*_{\nu}
\nonumber\\
& =& \frac{q \cdot P}{2} \varepsilon \cdot q \varepsilon'^* \cdot q
- \frac{q^2}{4} (\varepsilon \cdot P \varepsilon'^* \cdot q
- \varepsilon'^* \cdot P \varepsilon \cdot q')
- \frac {q \cdot q'}{2} \varepsilon'^* \cdot P \varepsilon \cdot q
\nonumber \\
& & 
- \frac{M}{2} \varepsilon \cdot q \varepsilon'^* \cdot q Q \cdot \gamma
+ M q \cdot q' \varepsilon \cdot q \varepsilon'^* \cdot \gamma
- \frac{M}{2} q^2 (\varepsilon \cdot q' \varepsilon'^* \cdot \gamma
- \varepsilon'^* \cdot q \varepsilon \cdot \gamma)
\nonumber \\
& & 
+ \frac{i}{4} q^2 (\varepsilon \cdot q' \sigma^{\nu \alpha}
\varepsilon'^*_{\nu} Q_{\alpha}
- \varepsilon'^* \cdot q \sigma^{\mu\alpha} \varepsilon_{\mu} Q_{\alpha})
+ \frac{i}{2} q \cdot q' q^2
\sigma^{\mu\nu} \varepsilon_{\mu} \varepsilon'^*_{\nu} \, ,
\\ \label{gl2_8i}
  \varepsilon_{\mu} \rho_{9}^{\mu \nu} \varepsilon'^*_{\nu}
& = & \varepsilon_{\mu} \tilde{T}_{14}^{\mu \nu} \varepsilon'^*_{\nu}
\nonumber\\
& = & \frac{q \cdot P}{2} ( \varepsilon \cdot P \varepsilon'^* \cdot q
- \varepsilon'^* \cdot P \varepsilon \cdot q')
- M q \cdot q' ( \varepsilon \cdot P \varepsilon'^* \cdot \gamma
- \varepsilon'^* \cdot P \varepsilon \cdot \gamma)
\nonumber \\
& & + M q \cdot P ( \varepsilon \cdot q' \varepsilon'^* \cdot \gamma
- \varepsilon'^* \cdot q \varepsilon \cdot \gamma)
+ \frac{i}{2} q \cdot q' ( \varepsilon \cdot P \sigma^{\nu\alpha}
\varepsilon'^*_{\nu} Q_{\alpha}
+ \varepsilon'^* \cdot P \sigma^{\mu \alpha} \varepsilon_{\mu} Q_{\alpha} )
\nonumber \\
& & - \frac{i}{2} q \cdot P (\varepsilon \cdot q' \sigma^{\nu \alpha}
\varepsilon'^*_{\nu} Q_{\alpha}
+ \varepsilon'^* \cdot q \sigma^{\mu \alpha} \varepsilon_{\mu} Q_{\alpha}) \, ,
\\ \label{gl2_8j}
  \varepsilon_{\mu} \rho_{10}^{\mu \nu} \varepsilon'^*_{\nu}
& = & \varepsilon_{\mu} \tilde{T}_{17}^{\mu \nu} \varepsilon'^*_{\nu}
\nonumber\\
& = & - 2 q \cdot P \varepsilon \cdot \varepsilon'^*
+ \varepsilon \cdot P \varepsilon'^* \cdot q
+ \varepsilon'^* \cdot P \varepsilon \cdot q'
+ 2 M \varepsilon \cdot \varepsilon'^* Q \cdot \gamma
\nonumber \\
& & 
- 2 M ( \varepsilon \cdot q' \varepsilon'^* \cdot \gamma
+ \varepsilon'^* \cdot q \varepsilon \cdot \gamma)
- i \varepsilon \cdot q' \sigma^{\nu \alpha}
\varepsilon'^*_{\nu} Q_{\alpha}
+ i \varepsilon'^* \cdot q \sigma^{\mu \alpha} \varepsilon_{\mu} Q_{\alpha}
\nonumber \\
& & 
- 2 i q \cdot q' \sigma^{\mu \nu} \varepsilon_{\mu} \varepsilon'^*_{\nu} \, ,
\\ \label{gl2_8l}
  \varepsilon_{\mu} \rho_{11}^{\mu \nu} \varepsilon'^*_{\nu}
& = & \varepsilon_{\mu} \tilde{T}_{18}^{\mu \nu} \varepsilon'^*_{\nu}
\nonumber\\
& = & (\varepsilon \cdot P \varepsilon'^* \cdot q
+ \varepsilon'^* \cdot P \varepsilon \cdot q') Q \cdot \gamma
- 2 q \cdot P (\varepsilon \cdot q' \varepsilon'^* \cdot \gamma
+ \varepsilon'^* \cdot q \varepsilon \cdot \gamma)
\nonumber \\
& & 
+ 2 i q \cdot q' \gamma_5 \varepsilon^{\mu\nu\alpha\beta}
\varepsilon_{\mu} \varepsilon'^*_{\nu} Q_{\alpha} \gamma_{\beta}
\, ,
\\ \label{gl2_8m}
  \varepsilon_{\mu} \rho_{12}^{\mu \nu} \varepsilon'^*_{\nu}
& = & \varepsilon_{\mu} \tilde{T}_{20}^{\mu \nu} \varepsilon'^*_{\nu}
=  \varepsilon_{\mu} \tilde{T}_{21}^{\mu \nu} \varepsilon'^*_{\nu}
\nonumber \\
& = & - \frac{q^{2}}{2} \varepsilon \cdot P \varepsilon'^* \cdot P
+ \frac{q \cdot P}{2} \varepsilon'^* \cdot P \varepsilon \cdot q
+ M q^{2} \varepsilon \cdot P \varepsilon'^* \cdot \gamma
- M q \cdot P \varepsilon \cdot q \varepsilon'^* \cdot \gamma
\nonumber \\
& & - \frac{i}{4} q^2 ( \varepsilon \cdot P \sigma^{\nu\alpha}
\varepsilon'^*_{\nu} Q_{\alpha}
+ \varepsilon'^* \cdot P \sigma^{\mu \alpha} \varepsilon_{\mu} Q_{\alpha} )
+ \frac{i}{2} q \cdot P \varepsilon \cdot q \sigma^{\nu \alpha}
\varepsilon'^*_{\nu} Q_{\alpha}
+ \frac{i}{2} q^{2} q \cdot P \sigma^{\mu \nu}
\varepsilon_{\mu} \varepsilon'^*_{\nu}
\nonumber \\
& & + \frac{i}{2} M q^{2} \gamma_5 \varepsilon^{\mu\nu\alpha\beta}
\varepsilon_{\mu} \varepsilon'^*_{\nu} Q_{\alpha} \gamma_{\beta}
\, .
\end{eqnarray}
\end{mathletters}
The sign of the Levi-Civit$\grave{\mbox{a}}$ symbol is fixed by
$\varepsilon_{0123} = - \varepsilon^{0123} = 1$, and
$\sigma_{\mu\nu} = i [ \gamma_{\mu} , \gamma_{\nu} ] / 2$
is the usual abbreviation for the commutator of the Dirac matrices.
Because of parity conservation Eqs. (\ref{gl2_8a})--(\ref{gl2_8m})
do not contain any pseudoscalar structures.

In analogy with the tensors, one can replace the
$B_{i} (q^{2} , 0 , q \cdot q' , q \cdot P)$ by 12 amplitudes
$f_{i} = f_{i}(q^{2} , q \cdot q' , q \cdot P)$,
\begin{eqnarray} \label{gl2_9}
&&
f_{1}=B_{1},\,\, 
f_{2}=B_{2},\,\,
f_{3}=B_{4},\,\,
f_{4}=B_{7},\,\,
f_{5}=B_{8}-B_{9},\,\,
f_{6}=B_{10},\nonumber\\
&&
f_{7}=B_{11},\,\,
f_{8}=B_{12} + B_{13},\,\,
f_{9}=B_{14},\,\,
f_{10}=B_{17},\,\,
f_{11}=B_{18},\,\,
f_{12}=B_{20} + B_{21}.
\end{eqnarray}

\section{Amplitude sets in virtual Compton scattering}
\label{appendix_b}
   Throughout this work we have applied the set of amplitudes defined in
Eqs. (\ref{gl2_16}) and (\ref{gl2_17}).
   The relation to the convention of Ref.\ \cite{Guichon_95} is given 
by 
\begin{eqnarray} \label{glb_3}
A_{1} & = & a^{t} \, ,
\nonumber \\
A_{2} & = & a^{t \prime} \, ,
\nonumber \\
A_{3} & = & - \sin \theta \, b_{1}^{t}
+ \sin \theta \cos \theta \, b_{1}^{t \prime}
- \sin \theta \, b_{2}^{t \prime}
- \sin^{2} \! \theta \, b_{3}^{t \prime} \, ,
\nonumber \\
A_{4} & = & - \frac{1}{\sin \theta} \, b_{2}^{t} \, ,
\nonumber \\
A_{5} & = & \frac{1}{\sin \theta} ( - \cos \theta \, b_{1}^{t}
+ b_{1}^{t \prime}
- \cos \theta \, b_{2}^{t \prime} ) \, ,
\nonumber \\
A_{6} & = & \frac{1}{\sin \theta} ( b_{1}^{t}
- \cos \theta \, b_{1}^{t \prime}
+ b_{2}^{t \prime} ) \, ,
\nonumber \\
A_{7} & = & \frac{1}{\sin \theta} ( - \cos \theta \, b_{1}^{t}
+ \cos^{2} \! \theta \, b_{1}^{t \prime}
- \cos \theta \, b_{2}^{t \prime}
+ \sin \theta \, b_{3}^{t}
- \sin \theta \cos \theta \, b_{3}^{t \prime}) \, ,
\nonumber \\
A_{8} & = & \frac{1}{\sin \theta} ( \cos^{2} \! \theta \, b_{1}^{t}
- \cos \theta \, b_{1}^{t \prime}
+ b_{2}^{t \prime}
- \sin \theta \cos \theta \, b_{3}^{t}
+ \sin \theta \, b_{3}^{t \prime}) \, ,
\nonumber \\
A_{9} & = & a^{l} \, ,
\nonumber \\
A_{10} & = & \frac{1}{\sin \theta} ( \cos \theta \, b_{1}^{l}
- b_{2}^{l}
- \sin \theta \, b_{3}^{l} ) \, ,
\nonumber \\
A_{11} & = & \sin \theta \, b_{1}^{l}
+ \cos \theta b_{3}^{l} \, ,
\nonumber \\
A_{12} & = & - b_{3}^{l} \, .
\end{eqnarray}


\end{document}